\documentclass[twocolumn,showpacs,preprintnumbers,amsmath,amssymb,superscriptaddress]{revtex4}

\usepackage{graphicx}
\usepackage{dcolumn}
\usepackage{bm}

\def\bea{\begin{eqnarray}}
\def\eea{\end{eqnarray}}

\begin{document}


\title{Event-wise $\langle p_t \rangle$ fluctuations
in Au-Au collisions at $\sqrt{s_{NN}}$ = 130 GeV}

%

\affiliation{Argonne National Laboratory, Argonne, Illinois 60439}
\affiliation{Brookhaven National Laboratory, Upton, New York 11973}
\affiliation{University of Birmingham, Birmingham, United Kingdom}
\affiliation{University of California, Berkeley, California 94720}
\affiliation{University of California, Davis, California 95616}
\affiliation{University of California, Los Angeles, California 90095}
\affiliation{Carnegie Mellon University, Pittsburgh, Pennsylvania 15213}
\affiliation{Creighton University, Omaha, Nebraska 68178}
\affiliation{Nuclear Physics Institute AS CR, \v{R}e\v{z}/Prague, Czech Republic}
\affiliation{Laboratory for High Energy (JINR), Dubna, Russia}
\affiliation{Particle Physics Laboratory (JINR), Dubna, Russia}
\affiliation{University of Frankfurt, Frankfurt, Germany}
\affiliation{Indiana University, Bloomington, Indiana 47408}
\affiliation{Insitute  of Physics, Bhubaneswar 751005, India}
\affiliation{Institut de Recherches Subatomiques, Strasbourg, France}
\affiliation{University of Jammu, Jammu 180001, India}
\affiliation{Kent State University, Kent, Ohio 44242}
\affiliation{Lawrence Berkeley National Laboratory, Berkeley, California 94720}\affiliation{Max-Planck-Institut f\"ur Physik, Munich, Germany}
\affiliation{Michigan State University, East Lansing, Michigan 48824}
\affiliation{Moscow Engineering Physics Institute, Moscow Russia}
\affiliation{City College of New York, New York City, New York 10031}
\affiliation{NIKHEF, Amsterdam, The Netherlands}
\affiliation{Ohio State University, Columbus, Ohio 43210}
\affiliation{Panjab University, Chandigarh 160014, India}
\affiliation{Pennsylvania State University, University Park, Pennsylvania 16802}\affiliation{Institute of High Energy Physics, Protvino, Russia}
\affiliation{Purdue University, West Lafayette, Indiana 47907}
\affiliation{University of Rajasthan, Jaipur 302004, India}
\affiliation{Rice University, Houston, Texas 77251}
\affiliation{Universidade de Sao Paulo, Sao Paulo, Brazil}
\affiliation{University of Science \& Technology of China, Anhui 230027, China}
\affiliation{Shanghai Institute of Nuclear Research, Shanghai 201800, P.R. China}
\affiliation{SUBATECH, Nantes, France}
\affiliation{Texas A\&M, College Station, Texas 77843}
\affiliation{University of Texas, Austin, Texas 78712}
\affiliation{Valparaiso University, Valparaiso, Indiana 46383}
\affiliation{Variable Energy Cyclotron Centre, Kolkata 700064, India}
\affiliation{Warsaw University of Technology, Warsaw, Poland}
\affiliation{University of Washington, Seattle, Washington 98195}
\affiliation{Wayne State University, Detroit, Michigan 48201}
\affiliation{Institute of Particle Physics, CCNU (HZNU), Wuhan, 430079 China}
\affiliation{Yale University, New Haven, Connecticut 06520}
\affiliation{University of Zagreb, Zagreb, HR-10002, Croatia}
\author{J.~Adams}\affiliation{University of Birmingham, Birmingham, United Kingdom}
\author{C.~Adler}\affiliation{University of Frankfurt, Frankfurt, Germany}
\author{M.M.~Aggarwal}\affiliation{Panjab University, Chandigarh 160014, India}
\author{Z.~Ahammed}\affiliation{Purdue University, West Lafayette, Indiana 47907}
\author{J.~Amonett}\affiliation{Kent State University, Kent, Ohio 44242}
\author{B.D.~Anderson}\affiliation{Kent State University, Kent, Ohio 44242}
\author{M.~Anderson}\affiliation{University of California, Davis, California 95616}
\author{D.~Arkhipkin}\affiliation{Particle Physics Laboratory (JINR), Dubna, Russia}
\author{G.S.~Averichev}\affiliation{Laboratory for High Energy (JINR), Dubna, Russia}
\author{S.K.~Badyal}\affiliation{University of Jammu, Jammu 180001, India}
% ...

\collaboration{STAR Collaboration}\noaffiliation

\date{\today}

\begin{abstract}
We present the first large-acceptance measurement of event-wise mean transverse momentum $\langle p_t \rangle$ fluctuations for Au-Au collisions at nucleon-nucleon center-of-momentum collision energy $\sqrt{s_{NN}} $ =  130 GeV. The observed non-statistical $\langle p_t \rangle$ fluctuations substantially exceed in magnitude fluctuations expected from the finite number of particles produced in a typical collision. The $r.m.s.$ fractional width excess of the event-wise $\langle p_t \rangle$ distribution is $13.7 \pm 0.1$(stat) $\pm 1.3$(syst)\% relative to a statistical reference, for the 15\% most-central collisions and for charged hadrons within pseudorapidity range $|\eta|<1$, $2\pi$ azimuth and $0.15 \leq p_t \leq 2$~GeV/$c$. The width excess varies smoothly but non-monotonically with collision centrality, and does not display rapid changes with centrality which might indicate the presence of critical fluctuations. The reported $\langle p_t \rangle$ fluctuation excess is qualitatively larger than those observed at lower energies and differs markedly from theoretical expectations. Contributions to $\langle p_t \rangle$ fluctuations from semi-hard parton scattering in the initial state and dissipation in the bulk colored medium are discussed.
\end{abstract}


\maketitle

\section{Introduction}
\label{Section:Intro}

Fluctuation analysis of relativistic heavy ion collisions has been
advocated to search for critical phenomena near the predicted hadron-parton phase boundary of quantum chromodynamics (QCD)~\cite{stock,tricrit,poly}. Nonstatistical fluctuations (excess variance beyond statistical fluctuations due to finite particle number), varying rapidly with collision energy, projectile size or collision centrality and interpreted as critical fluctuations, could indicate a transition to a quark-gluon plasma~\cite{stock,tricrit,poly}.  Nonstatistical fluctuations could also appear in systems incompletely equilibrated following initial-state multiple scattering (Cronin effect~\cite{iss} and minimum-bias hard parton scattering -- minijets~\cite{jetquench}), or as an aspect of fragmentation of color strings produced in nucleon-nucleon collisions or the bulk medium in A-A collisions. 
Study of nonstatistical fluctuations and the correlations which produce them are central aspects of the Relativistic Heavy Ion Collider (RHIC) research program.
The specific goal of the present work is to determine the magnitude and
collision centrality dependence of nonstatistical fluctuations in momentum
space at large momentum scales using the largest
angular acceptance detector available at RHIC.

The dynamical representation of relativistic nuclear collisions can be separated into transverse (perpendicular to the beam axis) and longitudinal (parallel to the beam axis) phase spaces.  In this paper we focus on transverse phase space, including transverse momentum magnitude $p_t$ and momentum azimuth angle $\phi$, within relatively small pseudorapidity $\eta$ intervals. Assuming rapid longitudinal (Bjorken) expansion of the collision system~\cite{schukraft}, separate $\eta$ intervals can be treated as  quasi-independent (causally disconnected) dynamical systems.  In this analysis we calculate the {\em event-wise mean transverse momentum} for each collision event within a detector kinematic acceptance
\bea
\langle p_t \rangle  \equiv \frac{1}{N}  \sum_{i=1}^{N} p_{t,i},
\label{Eq1}
\eea
where $i$ is a particle index and $N$ represents the measured
charged-particle multiplicity within the detector acceptance for a given collision event. Quantity $\langle p_t \rangle$ is monotonically related to the `temperature' of the event-wise $p_t$ distribution, plus any collective transverse velocity of the collision system.  The distribution of $\langle p_t \rangle$ over a collision event ensemble, especially any excess variance of this distribution beyond what is expected for purely statistical fluctuations, reflects the underlying dynamics and degree of equilibration of heavy ion collisions. 

Some aspects of heavy ion collisions produce correlations/fluctuations which depend on the relative charge of a charged hadron pair~\cite{axialcd,phenixq,unbalanced}, including quantum and Coulomb correlations~\cite{starhbt}, resonance decays, color-string fragmentation ({\em e.g.}, charge ordering along the string axis~\cite{lund,isrpp}), and minijet fragmentation.
Charge dependent combinations for pion pairs can be directly related to isospin
components.  For non-identified charged hadron pairs in the
collisions studied here, which are dominated by pions but include other charged
hadrons ({\em e.g.} protons, kaons and their antiparticles),
the relation to isospin
remains useful but becomes approximate.
To isolate the different isospin aspects of fluctuations and correlations we measure separately the like-sign (LS) and unlike-sign (US) charge-pair contributions and also form charge-independent (CI) and charge-dependent (CD) combinations, with CI = LS + US (approximately isoscalar) and CD = LS $-$ US (approximately isovector) respectively.

In this paper we report the first large-acceptance measurement of
$\langle p_t \rangle$ fluctuations at RHIC using the STAR detector. Results
are presented for unidentified charged hadrons using 183k central and 205k
minimum-trigger-bias ensembles of Au-Au collision events at
$\sqrt{s_{NN}} $ = 130 GeV (CM energy per nucleon-nucleon pair). Experimental details and the observed $\langle p_t \rangle$ distribution for central events are presented in Secs.~\ref{Section:Exp} and \ref{Section:Dist}.  Quantities used to measure nonstatistical $\langle p_t \rangle$ fluctuations are discussed in Sec.~\ref{Section:Measure} and the Appendix. Results and discussion are presented in Secs.~\ref{Section:Results} - \ref{Section:Disc}: the observed large excess of $\langle p_t \rangle$ fluctuations at RHIC is compared to other measurements and to theoretical models, including hard parton scattering in the initial state and/or hadronic rescattering.  Conclusions are presented in Sec.~\ref{Section:Con}.

\section{Summary of Experiment}
\label{Section:Exp}

Data for this analysis were obtained with the STAR detector~\cite{star} employing a 0.25~T uniform magnetic field parallel to the beam axis. Event triggering with the central trigger barrel (CTB) scintillators and zero-degree calorimeters (ZDC) and charged-particle kinematic measurements with the time projection chamber (TPC) are described in~\cite{star}. TPC tracking efficiency was determined to be 80 - 95\% within $|\eta| < 1$ and $p_t  > 200$ MeV/$c$ by embedding simulated tracks in real-data events~\cite{spectra}, and was uniform in azimuth to 3\% ($r.m.s.$) over 2$\pi$. Split-track removal required the fraction of valid space points used in a track fit relative to the maximum number possible to be $> 50\%$.  A primary event vertex within 75~cm of the axial center of the TPC was required. Valid TPC tracks fell within the full detector acceptance, defined here by $0.15 < p_t < 2.0$~GeV/$c$, $|\eta| < 1$ and $2 \pi$ in azimuth. Primary tracks were defined as having a distance of closest approach less than 3~cm from the reconstructed primary vertex which included a large fraction of true primary hadrons plus approximately 7\% background contamination~\cite{spectra}.

Two data sets were analyzed: 1) 183k central triggered Au-Au
collision events constituting the 15\% most-central collisions as determined
by scintillator hits in the STAR CTB and 2) 205k minimum-bias collision events triggered by ZDC coincidence.  The latter events were divided into eight centrality classes based on TPC track multiplicity in $|\eta| \leq 0.5$~\cite{spectra}, the eight event classes comprising approximately equal fractions of the upper 87$\pm$2\% of the Au-Au total hadronic cross section.


\section{Mean $p_t$ distribution}
\label{Section:Dist}

\begin{figure}[h]
\includegraphics[keepaspectratio,width=3.3in]{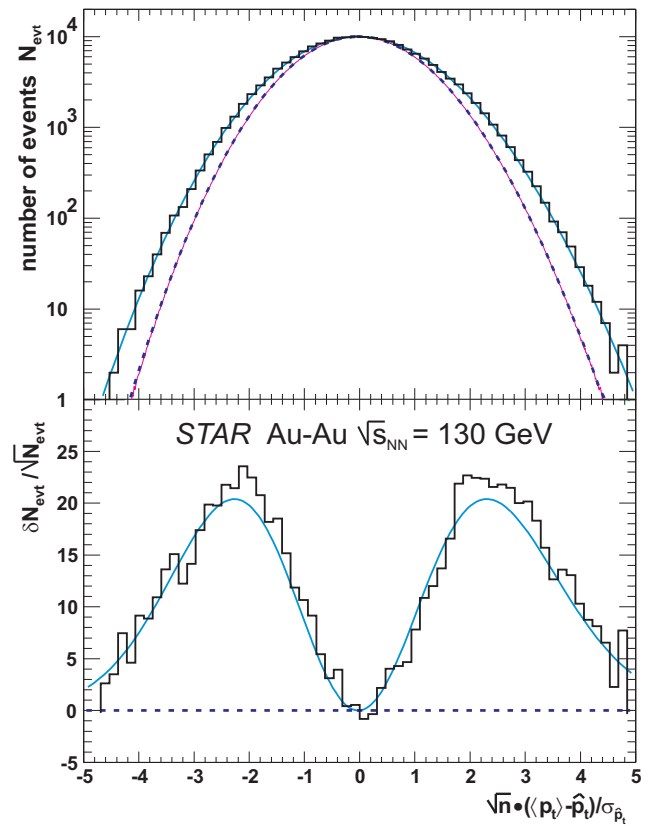}
\caption{\label{pt1d} Upper panel:
Event frequency distribution on $ \sqrt{n}\, (\langle p_t \rangle - \hat p_t)
/ \sigma_{\hat p_t}$ (see text) for 80\% of primary charged hadrons in $|\eta|<1$ for 183k central events (histogram) compared to gamma reference (dashed curve), Monte Carlo reference (solid curve underlying gamma reference), and broadened
distribution (solid curve underlying data, not a fit -- see text). Lower panel: Difference in upper panel between data
and gamma reference (histogram) or between broadened distribution and gamma reference (solid curve) normalized by the Poisson error $\sqrt{N_{evt}}$ in each bin.
}
\end{figure}

The frequency distribution of event-wise $\langle p_t \rangle$ for 183k 15\% most-central collision events is first studied graphically. The data histogram is compared to a statistical reference distribution and is examined for evidence of anomalous event classes which could indicate either novel collision dynamics~\cite{stock} or experimental anomalies. The event-wise $\langle p_t \rangle$ data distribution is shown as the histogram in the upper panel of Fig.~\ref{pt1d}.  Those data, representing $80 \pm 5$\% of the
true primary particles within the acceptance, were binned using quantity $(\langle p_{t} \rangle - \hat p_t) / (\sigma_{\hat p_t}/\sqrt{n})$,  where $\hat{p}_t$ and $\sigma^2_{\hat p_t}$ are respectively the mean and variance of the inclusive $p_t$ distribution of all accepted particles in  the event ensemble and $n$ is the event-wise multiplicity within the defined acceptance. That choice of event-wise {\em random variable} rather than $\langle p_{t} \rangle$ is explained as follows.

For independent particle $p_t$ samples from a fixed {\em parent} distribution (no nonstatistical fluctuations) the {\em r.m.s.} width of the frequency distribution on $\langle p_{t} \rangle$ is itself dependent on event multiplicity $n$ as $\sigma_{\hat p_t}/\sqrt{n}$ (central limit theorem or CLT~\cite{handbook,cltps}). The underlying purpose of this measurement is to determine an aspect of $p_t$ fluctuations which is {\em independent of}\, event multiplicity {\em per se}. If $n$ is a random variable, a systematic dependence is introduced into the measured $\langle p_{t} \rangle$ fluctuation excess through this CLT behavior of the width. To insure multiplicity independence the basic statistical quantity must be formulated carefully. By normalizing the distribution variable with factor $\sqrt{n} / \sigma_{\hat p_t}$, the distribution width of the new variable is unity, independent of $n$, when fluctuations are purely statistical.  The trivial broadening of the $\langle p_t \rangle$ distribution for event ensembles with a finite range of event multiplicities is eliminated. The latter effect can have significant consequences for relevant event ensembles (p-p, peripheral A-A and small detector acceptance). This argument explains the variable choice for Fig.~\ref{pt1d} as well as the associated numerical analysis described in Sec.~\ref{Section:Measure}. For the sake of brevity this normalized variable will in some cases still be referred to in the text as ``$\langle p_{t} \rangle$''.

The precision of these data warrants construction of a statistical reference which accurately represents the expected $\langle p_t \rangle$ distribution in the absence of nonstatistical fluctuations. Because of its close connection to the central limit theorem (behavior under $n$-folding noted below) we can compactly and accurately represent the $\langle p_t \rangle$ reference distribution with a gamma distribution~\cite{tann}. We observe that the measured inclusive $p_t$ distribution is, for present purposes, well approximated by a gamma distribution with folding index $\alpha_0 \equiv {\hat p_t}^2 / \sigma^2_{\hat p_t} \approx 2$. Differences between the gamma and inclusive $p_t$ distributions in the higher cumulants due to $p_t$ acceptance cuts and physics correlations are strongly suppressed in the comparison with the distribution in Fig.~\ref{pt1d} by inverse powers of event multiplicity and are not significant for central Au-Au collisions. 


Because the $n$-folding of a gamma distribution is also a gamma distribution (representing an ensemble of independent $n$-samples of the parent gamma distribution or inclusive $p_t$ distribution), the $\langle p_t \rangle$ reference distribution can be represented by~\cite{tann}
\bea
g_{\bar n}(\langle p_t \rangle) =  \frac{\alpha_0 }{  \hat p_t} \cdot \,
\frac{ e^{-\alpha_0 \,
\bar n \langle p_t \rangle / \hat p_t} }{ \Gamma(\alpha_0 \, \bar n )} \cdot
\left(\alpha_0 \,  \bar n
\frac{\langle p_t \rangle }{  \hat p_t}\right)^{\alpha_0
 \,
\bar n  -1} . 
\label{Eq2}
\eea
The corresponding gamma-distribution reference is indicated by the dashed curve in the upper panel of Fig.~\ref{pt1d}.  Parameter values used for this reference curve were determined from the measured inclusive $p_t$ distribution as $\bar{n} = 735\pm0.2$, $\hat{p}_t = 535.32 \pm$0.05~MeV/$c$, and $\sigma_{\hat{p}_t} = 359.54 \pm$ 0.03~MeV/$c$, obtained from all accepted particles and not corrected for $p_t$ acceptance cuts and inefficiencies.

A reference can also be generated by a Monte Carlo procedure. An ensemble of $n$-sample reference events is generated with multiplicity distribution similar to the data.  A reference event with multiplicity $n$ drawn from that distribution is assembled by performing $n$ random samples from a fixed parent $p_t$ distribution estimated by the interpolated inclusive $p_t$ histogram of all accepted particles from all events in the centrality bin. The resulting Monte Carlo reference distribution is shown in Fig.~\ref{pt1d} upper panel by the solid curve underlying the dashed gamma reference curve. The agreement is excellent. The broadened distribution (solid curve) underlying the data in the upper panel of Fig.~\ref{pt1d} is discussed in Sec.~\ref{Section:Results}. All curves are normalized to match the data near the peak value, emphasizing the width comparison which is the main issue of this paper. We observe a substantial width excess in the data relative to the statistical reference.

Fig.~\ref{pt1d} lower panel shows the difference between data and gamma
reference normalized to Poisson standard deviations in each bin,
emphasizing the large statistical significance of the width excess.  We observe no significant deviations (bumps) from the broadened distribution in Fig.~\ref{pt1d} which might indicate anomalous event classes as expected in some phase-transition scenarios~\cite{stock}. It is also important to note that the entire event ensemble contributes to the width increase relative to the statistical reference, {\em i.e.,} the excess width is not dominated by a subset of problematic events. We note that the distribution in Fig.~\ref{pt1d} cannot be corrected for background contamination and tracking inefficiency. The numerical analysis described in the next section allows such corrections.

\section{Measures of nonstatistical $\langle p_t \rangle$ fluctuations}
\label{Section:Measure}

Consistent with the argument presented above about eliminating dependence of fluctuation measures on multiplicity variations within a centrality bin, we characterize the magnitude of nonstatistical $\langle p_t \rangle$ fluctuations by comparing the variance of distribution quantity $\sqrt{n} (\langle p_t \rangle - \hat p_t)$ from Fig.~\ref{pt1d} to the variance $\sigma^2_{\hat p_t}$ of its reference distribution. The difference between these two variances is represented by
\bea \label{Eq3}
\Delta\sigma^{2({\rm CI})}_{p_t:n} & \equiv & 
\frac{1}{\varepsilon} \sum_{j=1}^{\varepsilon} n_j \left[
\langle p_t \rangle_j - \hat{p}_t \right]^2 - \sigma^2_{\hat p_t} \\
 & \equiv & 2 \sigma_{\hat p_t} \Delta\sigma^{({\rm CI})}_{p_t:n},
\label{Eq4}
\eea
where $\varepsilon$ is the number of events in a centrality bin, $j$ is the event index, $n_j$ is the number of accepted particles in event $j$, $\langle p_t \rangle_j$ is the mean $p_t$ of accepted particles in event $j$. Subscript $p_t\! :\! n$ emphasizes that this quantity measures variance excess due to fluctuations of $p_t$ {\em relative to} event-wise fluctuations in multiplicity $n$ ({\em i.e.,} is not significantly affected by fluctuations in $n$ itself). Superscript (CI) indicates a {\em charge-independent} sum over all particles. {\em Difference factor} $\Delta\sigma^{({\rm CI})}_{p_t:n}$ defined in Eq.~(\ref{Eq4}) is approximately equal to $\langle p_t \rangle$ fluctuation measure $\Phi_{p_t}$ introduced previously~\cite{phipt,guntherphipt}.
 
Two issues motivate the definition of fluctuation measure $\Delta\sigma^{2({\rm CI})}_{p_t:n}$ in Eq.~(\ref{Eq3}): 1) $\langle p_t \rangle$ is the ratio of two random variables -- a scalar $p_t$ sum and a multiplicity.  Fluctuations in either variable contribute to fluctuations in the ratio. For an uncorrelated system with fluctuating multiplicity, ratio fluctuations go as $1/\sqrt{n}$, producing an apparent nonstatistical contribution to ratio fluctuation measures which are aimed at determining $p_t$ fluctuations.  2) Measures of nonstatistical fluctuations typically involve (at least implicitly) a difference between variances evaluated at two different {\em scales}, where `scale' in the present context refers to histogram bin sizes ({\em e.g.,} on $\eta$ and $\phi$). Bins on $\eta$ and $\phi$ are denoted respectively by $\delta\eta$ and $\delta\phi$ or generically by $\delta x$.  The detector acceptance can define one scale, as in this analysis.  
The other relevant scale, both for the simulated events presented in the
preceding section and in the variance measurements presented in Sec.~\ref{Section:Results},
is the {\em single particle} scale in which the bins are always made small
enough such that occupied bins contain a maximum of one particle.  In
general the scale is independent of the acceptance where
{\em scale} $\leq$ {\em acceptance}.  The case of variance calculations for
arbitrary scale is treated in the Appendix.  Scale dependence of variance excess provides important information on
the underlying two-particle correlations and is an {\em essential feature} of any nonstatistical fluctuation measurement such as those presented here,
although the importance of this point has not been fully appreciated in this heavy ion context.

In the Appendix we show that the {\em scale invariance of total variance}, an expression of the CLT, motivates the quantity in Eq.~(\ref{Eq3}). $\Delta\sigma^{2({\rm CI})}_{p_t:n}(\delta x)$ measures changes in variance stemming from two-particle correlations with characteristic lengths less than binning scale $\delta x$~\cite{cltps}. As a function of binning scale $\Delta\sigma^{2({\rm CI})}_{p_t:n}(\delta x)$ is not dependent on an acceptance {\em size} (knowledge of its scale dependence may of course be limited by a finite detector acceptance) but can depend on {\em absolute position} of the acceptance in momentum space.

Given the definition of $\Phi_{p_t}$~\cite{phipt} and Eq.~(\ref{Eq3}), $\Delta \sigma^{2({\rm CI})}_{p_t:n} \simeq (\Phi_{p_t} + \sigma_{\hat p_t})^2 - \sigma^2_{\hat p_t}$, and $\Phi_{p_t} \simeq \Delta \sigma^{({\rm CI})}_{p_t:n}$~\cite{cltps}. Difference factor $\Delta \sigma^{({\rm CI})}_{p_t:n}$ and $\Phi_{p_t}$ are therefore comparable between different analyses. Fluctuation measure $\sigma^2_{p_t,dyn}\equiv \overline{ \langle ( p_{t,i} - \hat{p}_t )( p_{t,j} - \hat{p}_t ) \rangle} _{i \neq j } $~\cite{sergei} (bar denotes event average) is related to $\Delta \sigma^{2({\rm CI})}_{p_t:n}$ by
$\sigma^2_{p_t,dyn} \simeq
\Delta \sigma^{2({\rm CI})}_{p_t:n} / (\bar N - 1)$ ($\bar{N}$ is the mean
multiplicity) for approximately constant event-wise multiplicities.
$\Phi_{p_t}$ and $\sigma^2_{p_t,dyn}$ may include
significant dependence on multiplicity fluctuations
in the case of small bin multiplicities ({\em e.g.}, for any bins within p-p or peripheral A-A events or for small-scale bins within central A-A events).
Variance difference $\Delta \sigma^{2({\rm CI})}_{p_t:n}$ minimizes this dependence compared to the preceding quantities.

In Eqs.~(\ref{Eq3}), (\ref{Eq4}) and the Appendix the summations over particles have ignored charge sign. $\Delta \sigma^{2({\rm CI})}_{p_t:n}$ is a {\em charge-independent} (approximately isoscalar) quantity. By separating contributions to Eq.~(\ref{Eq3}) into sums over $(+)$ and $(-)$ charges, charge-dependent (CD) quantity $\Delta \sigma^{2({\rm CD})}_{p_t:n}$ can be defined which measures the difference between contributions to $\langle p_t \rangle$ fluctuations from like-sign pairs and unlike-sign pairs.  Using explicit
charge-sign notation quantities $\Delta \sigma^{2({\rm CI})}_{p_t:n}$ and
$\Delta \sigma^{2({\rm CD})}_{p_t:n}$ are defined by
\bea \label{EqCI}
\bar{N}(\Delta x)\, \Delta \sigma^{2({\rm CI})}_{p_t:n}  &=&  
     \bar{N}(\Delta x)_{+} \Delta \sigma^2_{p_t:n,++} \\ \nonumber
  &+&  \bar{N}(\Delta x)_{-} \Delta \sigma^2_{p_t:n,--}   \\ \nonumber
    &+& 2 \sqrt{ \bar{N}(\Delta x)_{+} \bar{N}(\Delta x)_{-} }
          \Delta \sigma^2_{p_t:n,+-}   
\eea
\bea \label{EqCD}
\bar{N}(\Delta x) \,\Delta \sigma^{2({\rm CD})}_{p_t:n}  &=&  
     \bar{N}(\Delta x)_{+} \Delta \sigma^2_{p_t:n,++} \\ \nonumber
  &+&  \bar{N}(\Delta x)_{-} \Delta \sigma^2_{p_t:n,--}  \\ \nonumber
    &-& 2 \sqrt{ \bar{N}(\Delta x)_{+} \bar{N}(\Delta x)_{-} }
          \Delta \sigma^2_{p_t:n,+-}
\eea
where $\bar{N}(\Delta x)_{\pm}$ are the mean multiplicities for $\pm$
charges in acceptance $\Delta x$ and $\bar{N}(\Delta x)$ is the mean total
multiplicity in $\Delta x$.  Individual terms in Eqs.~(\ref{EqCI}) and~(\ref{EqCD})  are defined by
\bea
\Delta \sigma^2_{p_t:n,ab}   & \equiv & 
\overline{\sqrt{n_a} \left( 
\langle p_t \rangle_a - \hat{p}_{ta} \right) \sqrt{n_b} \left(
\langle p_t \rangle_b - \hat{p}_{tb} \right) } \nonumber \\
\label{EqSig-ab}
 & &  - \sigma^2_{\hat{p}_{t,a}} \delta_{ab},
\eea
where subscripts $a$ and $b$ represent charge sign,
$ab = ++, --, +-$ or $-+$, the bar denotes an average over events, and
$\delta_{ab}$ is a Kronecker delta. 
Difference factors
$\Delta \sigma^{({\rm CI})}_{p_t:n}$ and $\Delta \sigma^{({\rm CD})}_{p_t:n}$ (approximately isoscalar and isovector respectively) reported in the following sections are defined by 
\bea
\label{EqCIfactor}
\Delta \sigma^{2({\rm CI})}_{p_t:n} & = & 2 \sigma_{\hat{p}_t}
\Delta \sigma^{({\rm CI})}_{p_t:n} \\
\label{EqCDfactor}
\Delta \sigma^{2({\rm CD})}_{p_t:n} & = & 2 \sigma_{\hat{p}_t}
\Delta \sigma^{({\rm CD})}_{p_t:n}.
\eea

\section{Results}
\label{Section:Results}

\begin{figure}[h]
\includegraphics[keepaspectratio,width=3.3in]{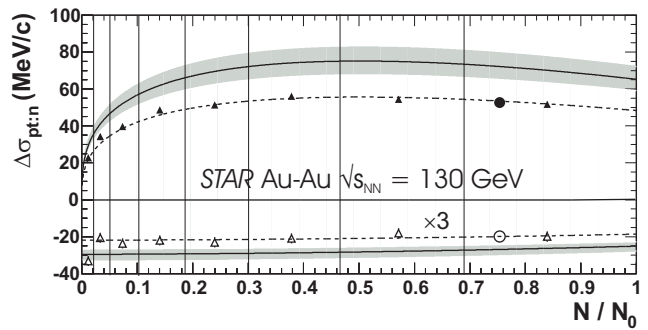}
\caption{\label{minbias} Mean-$p_t$ difference factors
$\Delta\sigma^{({\rm CI})}_{p_t:n}$ and $\Delta\sigma^{({\rm CD})}_{p_t:n}$ for 205k
minimum-bias Au-Au events at $\sqrt{s_{NN}}$ = 130 GeV versus relative
multiplicity $N/N_0$~\cite{spectra}, which is approximately
$N_{part}/N_{part,max}$, the relative fraction of participant nucleons~\cite{endpoint}.
Charge-independent (solid triangular points) and
charge-dependent (open triangular points -
multiplied by 3 for clarity) difference
factors include statistical errors only (smaller than symbols).
Parametrizations (dashed curves), extrapolation of parametrizations to true
primary particle
number (solid curves), and systematic uncertainties (bands) are discussed in
the text.  Difference factors for the 15\% most-central collision events are
shown by the solid circle and open circle symbols.
}
\end{figure}

We apply Eqs.~(\ref{EqCI}) -- (\ref{EqCDfactor}) to
central collisions and to a minimum-bias ensemble.
In all cases charge symmetry 
$\Delta \sigma^2_{p_t:n,++} \simeq \Delta \sigma^2_{p_t:n,--}$
is observed within errors.
For the 15\% most-central events and full acceptance we obtain difference factors $\Delta\sigma^{({\rm CI})}_{p_t:n} = 52.6\pm0.3$ (stat)~MeV/$c$ and
$\Delta\sigma^{({\rm CD})}_{p_t:n} = -6.6\pm0.6$ (stat)~MeV/$c$ (respectively the solid and open circular data symbols in Fig.~\ref{minbias}). Charge-independent values of $\Phi_{p_t}$ and $\sigma^2_{p_t,dyn}$ for the same data are respectively $52.6 \pm 0.3$ (stat)~MeV/$c$ and $52.3 \pm 0.3$ (stat)~(MeV/$c$)$^2$ (note units).
Dependence on multiplicity fluctuations is negligible for this full-acceptance 15\% most-central collision ensemble. 

The experimental value $\Delta\sigma^{({\rm CI})}_{p_t:n} = 52.6$ MeV/$c$ was used to determine the solid curves underlying the data histogram in the two panels of Fig.~\ref{pt1d} by raising the reference gamma distribution 
in Eq.~(\ref{Eq2}) to the power $\sigma^2_{\hat p_t}/(\sigma^2_{\hat p_t} +
\Delta\sigma^{2({\rm CI})}_{p_t:n})$. This procedure, which would be exact for a gaussian distribution, increases the variance of the modified gamma distribution to the numerical value obtained from the data, preserves the mean, and agrees well with the relative peak heights of the data in the lower half of Fig.~\ref{pt1d}. The comparison in Fig.~\ref{pt1d} then demonstrates that $\Delta\sigma^{({\rm CI})}_{p_t:n}$ provides an excellent description of the event-wise $\langle p_t \rangle$ distribution and its fluctuation excess. The corresponding $r.m.s.$ width increase relative to the reference is $13.7\pm0.1$(stat)$\pm 1.3$(syst)\%. When extrapolated to 100\% of primary hadrons and no backgrounds $\Delta\sigma^{({\rm CI,CD})}_{p_t:n}$ was estimated to be a factor 1.26 larger in magnitude for the 15\% most-central events, resulting in a corrected charge-independent {\em r.m.s.} width increase of $17 \pm 2$(syst)\%.

Difference factors were also determined for eight centrality classes defined for the 205k minimum-bias events described in Sec.~\ref{Section:Exp}. Measured values of $\Delta\sigma^{({\rm CI})}_{p_t:n}$ and $\Delta\sigma^{({\rm CD})}_{p_t:n}$  are shown in Fig.~\ref{minbias} by the upper and lower set of data symbols for CI and CD respectively, plotted for each centrality class {\em vs} its mean multiplicity $\bar{N}$ in $|\eta| \leq 0.5$ (Sec.~\ref{Section:Exp}) relative to $N_0$, the minimum-bias multiplicity distribution endpoint~\cite{endpoint} where $N_0 = 520 \pm 5$. Data are listed in Table~\ref{TableI}.  Plotted points, including statistical errors only (typically $\pm 0.5$~MeV/$c$), were fitted with parametrizations (dashed curves) which were then extrapolated by amounts varying from 1.17 to 1.26 (for peripheral to central events respectively) to produce estimates for 100\% of primary charged hadrons (solid curves). $\Delta\sigma^{({\rm CI})}_{p_t:n}$ has a very significant non-monotonic dependence on centrality, but with no sharp structure. $\Delta\sigma^{({\rm CD})}_{p_t:n}$ is significantly negative and approximately independent of centrality. 
$\Phi_{p_t}$ and $\sigma^2_{p_t,dyn}(\bar{N} - 1)/2\sigma_{\hat p_t}$ agree with $\Delta\sigma^{({\rm CI})}_{p_t:n}$ within statistical errors for the upper six centrality classes, but both differ from $\Delta\sigma^{({\rm CI})}_{p_t:n}$ and each other by much more than statistical uncertainty for the two most peripheral bins, as expected from their dependencies on multiplicity fluctuations.

\begin{table}[h]
\caption{\label{TableI}
Centrality dependences of the measured charge independent (CI) and charge
dependent (CD) difference factors $\Delta\sigma^{({\rm CI})}_{p_t:n}$
 and $\Delta\sigma^{({\rm CD})}_{p_t:n}$ plus the corresponding
values extrapolated to
100\% tracking efficiency.
}
\begin{tabular}{|ccc|cc|cc|} \hline
\multicolumn{3}{|c|}{Centrality} & \multicolumn{2}{c|}{$\Delta\sigma^{({\rm CI})}_{p_t:n}$ (MeV/$c$)} & \multicolumn{2}{c|}{$\Delta\sigma^{({\rm CD})}_{p_t:n}$ (MeV/$c$)} \\ 
\hline
$\bar{N}/N_0$  &  $\frac{\sigma}{\sigma_{tot}}$(\%)\footnote{Fraction of total hadronic inelastic cross section ranges in percent; values are $\pm2$\% uncertain~\cite{spectra}.}  &  $N_{part}$\footnote{Estimates in \cite{spectra} were interpolated to centrality bins used here.}  & Data\footnote{Statistical errors are
typically $\pm$0.5~MeV/$c$; systematic errors are $\pm$9\%.}
& Ext.\footnote{Difference factors extrapolated to 100\% tracking efficiency and no secondary particle contamination.  Uncertainties are $\pm$12\% and are dominated by systematics as discussed in the text.}  & Data$^{c}$ & Ext.$^{d}$  \\ 
\hline
 0.012  &  87-76  &  8.9  &  22.8  &  26.8  & -11.1  &  -13.0 \\
 0.033  &  76-65  &  19   &  34.3  &  40.4  &  -6.9  &  -8.1  \\
 0.073  &  65-54  &  36   &  39.6  &  46.8  &  -7.9  &  -9.3  \\
 0.14   &  54-43  &  64   &  48.7  &  57.6  &  -7.4  &  -8.8  \\
 0.24   &  43-33  &  102  &  51.3  &  61.4  &  -7.7  &  -9.2  \\
 0.38   &  33-22  &  153  &  56.1  &  68.0  &  -7.0  &  -8.5  \\
 0.57   &  22-11  &  224  &  54.4  &  66.9  &  -6.0  &  -7.4  \\
 0.84   &  11-0   &  320  &  51.8  &  65.1  &  -6.6  &  -8.3  \\
\hline
\end{tabular}
\end{table}

Systematic errors from uncertainty in two-track inefficiency,
primary-vertex transverse position uncertainty,
TPC drift speed/time-offset uncertainty, and conversion electron contamination
were estimated by Monte Carlo~\cite{lanny} as less than 4\% of reported values.
Stability of reported results against
primary-vertex longitudinal position variation,
momentum resolution, and
TPC central membrane track crossing was determined to be 5\% of
stated values.
Systematic effects due to possible time dependence in detector performance and
efficiency were studied by analyzing sequential
run blocks which were determined to be
consistent within statistical error.  Systematic error contributions due to
azimuthal anisotropy in the event-wise primary particle distribution
($\cos [2(\phi - \Psi_R)]$ assumed where $\Psi_R$ is the event-wise reaction plane angle) combined with non-uniform azimuthal tracking efficiency
was determined to be less than 1\% of reported values using $\phi$-dependent
track cuts and measured efficiency maps.
Nonprimary background ($\sim$7\%)~\cite{spectra}
added $\pm7$\% systematic error due to
uncertainty in its correlation content.
Total systematic uncertainty for the $\Delta\sigma^{({\rm CI})}_{p_t:n}$
and $\Delta\sigma^{({\rm CD})}_{p_t:n}$
data in Fig.~\ref{minbias} and Table~\ref{TableI} is $\pm$9\%.
Additional systematic error in extrapolation of
$\Delta\sigma^{({\rm CI})}_{p_t:n}$
and $\Delta\sigma^{({\rm CD})}_{p_t:n}$
to 100\% of primary particles
($\pm$8\%) is dominated by uncertainty
in the actual primary particle yield \cite{spectra}.
Total uncertainty in extrapolated
values is about $\pm$12\% (shaded bands in
Fig.~\ref{minbias}).  Systematic error in the most peripheral bin is
larger by an additional $\sim\pm$1~MeV/$c$
due to possible primary-vertex reconstruction bias.
Analyses of 30k central {\sc hijing} Au-Au collision events both with and
without STAR acceptance and event reconstruction effects yield
consistent results for $\Delta\sigma^{({\rm CI})}_{p_t:n}$ to within 
the statistical error ($\sim$10\%) for these simulated events, which is well within our
estimated systematic error.

Data in Fig.~\ref{minbias} and Table~\ref{TableI} were not corrected for two-track inefficiencies, which would increase all results in a positive sense by up to 3~MeV/$c$.
Variations ($\approx 10\%$) in $\hat p_t$ and $\sigma^2_{\hat p_t}$
with collision centrality were accommodated by independent analyses in small
centrality bins. Monte Carlo~\cite{lanny} estimates indicate that combined
corrections for quantum (HBT) and Coulomb correlations~\cite{starhbt}, resonance ($\rho^0$, $\omega$) decays, and $\hat{p}_t$  centrality dependence ({\em i.e.,} well known physical effects) would {\em increase} the absolute magnitudes of all data in Fig.~\ref{minbias} and Table~\ref{TableI} by as much as $\approx 6$~MeV/$c$. Quantum and Coulomb correlations and resonance decays originate in the final stage of the collision evolution and are not the main object of this study.  Correcting $\Delta\sigma^{({\rm CI})}_{p_t:n}$ for two-track inefficiencies plus the preceding effects (not done for the data shown in Fig.~\ref{minbias} and Table~\ref{TableI}) would cause the overall magnitude to increase by
about 7~MeV/$c$.  Similarly, corrections to $\Delta\sigma^{({\rm CD})}_{p_t:n}$ would cause it to become more negative by about 4~MeV/$c$.  We conclude that the negative values of $\Delta\sigma^{({\rm CD})}_{p_t:n}$ are physically significant, and cannot be explained by conventional effects such as Coulomb interactions, resonance decays or tracking inefficiencies. 

\section{Experiment Comparisons}
\label{Section:ExComp}

CERN Super Proton Synchrotron (SPS) charge-independent $\Phi_{p_t}$
measurements with a 158~GeV per nucleon Pb beam on fixed heavy ion targets
($\sqrt{s_{NN}} \approx 17.3$ GeV) include values $0.6 \pm 1.0$~MeV/$c$
for central collisions on Pb nuclei with $\bar{N} \simeq 270$ in CM pion rapidity interval $1.1 \leq y_{\pi,cm} \leq 2.6$ (experiment NA49)~\cite{guntherphipt} and $3.3 \pm 0.7^{+1.8}_{-1.6}$~MeV/$c$ for central collisions on Au nuclei with $\bar{N} \simeq 162$ in laboratory pseudorapidity interval $2.2 \leq \eta_{lab} \leq 2.7$ (mid-pseudorapidity region) from the CERES experiment~\cite{ceres}. STAR measures $\Delta \sigma^{({\rm CI})}_{p_t:n} = 14 \pm 2$~MeV/$c$ $\simeq \Phi_{p_t}$ for $\bar{N} \sim 180$ when restricted to the CERES $\eta$ acceptance scale~\cite{ceres}. All three measurements were corrected for small-scale correlations and two-track inefficiencies.
In a following analysis~\cite{anticic} of the 158~GeV per nucleon Pb-Pb
fixed target collision data experiment NA49 reported charge independent
$\Phi_{p_t}$ measurements for all charged particles in rapidity interval
$1.1 \leq y_{\pi,cm} \leq 2.6$ (pion mass assumed) as a function of centrality.
$\Phi_{p_t}$ values were found to monotonically decrease from
$7.2 \pm 0.7 \pm 1.6$~MeV/$c$ for most-peripheral to
$1.4 \pm 0.8 \pm 1.6$~MeV/$c$ for most-central collisions.  Corrections for
finite two-track resolution were included, however the contributions of
quantum and Coulomb small-scale correlations, estimated to be
$5\pm 1.5$~MeV/$c$~\cite{guntherphipt}, remain.
Quantity $\Sigma_{p_t} \equiv \sqrt{\Delta\sigma^{2({\rm CI})}_{p_t:n}
/\bar{N} \hat{p}_t^2}$ was also reported by the CERES experiment~\cite{ceres}
with magnitude approximately half that at STAR. Results from STAR for $\Delta \sigma^{({\rm CI})}_{p_t:n}$ at RHIC energy represent a striking increase over SPS results and markedly different centrality dependence. In contrast, STAR's measurement of $\Delta \sigma^{({\rm CD})}_{p_t:n}$ is not significantly
different from the NA49 result $-8.5 ~\pm 1.5$ MeV/$c$ in $1.1 \leq y_{\pi,cm} \leq 2.6$~\cite{jeffdata}.

The PHENIX experiment at RHIC reports charge-independent 
$\Phi_{p_t} \approx 6\pm~6$~(syst)~MeV/$c$ for the upper-most 5\%
central Au-Au collision events at $\sqrt{s_{NN}} = 130$~GeV within their acceptance: $|\eta|< 0.35$ and $\Delta \phi = 58.5^{\circ}$~\cite{phenix}.  This STAR analysis restricted to the PHENIX $(\eta,\phi)$ acceptance scale obtained $\Delta \sigma^{({\rm CI})}_{p_t:n}\sim 9 \pm 1$~MeV/$c$. That value is greater than would be expected from naive scaling from the STAR full acceptance scale ($\Delta \eta = 2,~\Delta\phi = 2\pi$) to the PHENIX acceptance scale ($\Delta \eta = 0.7,~\Delta\phi = 58.5^{\circ}$)~\cite{footnote-scale}.  The enhanced value for $\Delta \sigma^{({\rm CI})}_{p_t:n}$ relative to linear scale dependence is observed to result from substantial nonlinear azimuth-scale ($\delta\phi$) dependence of $\langle p_t \rangle$ fluctuations (mainly a $\cos [2(\phi - \Psi_R)]$ term). 

PHENIX also reports non-zero nonstatistical $\langle p_t \rangle$ fluctuations for Au-Au collisions at $\sqrt{s_{NN}}$ = 200~GeV using quantity $F_{p_t}$~\cite{phenix200} (proportional to $\Phi_{p_t}$ and $\Delta\sigma^{({\rm CI})}_{p_t:n}$) and acceptance scales $\Delta \eta = 0.7$ at mid-rapidity and $\Delta\phi = 180^{\circ}$ in two approximately opposed $90^{\circ}$ spectrometer arms.  $\langle p_t \rangle$ fluctuations for central collisions at 200~GeV (with two opposed spectrometer arms) are observed to be similar to those at 130~GeV (with one spectrometer arm) assuming linear dependence on azimuth scale~\cite{footnote-scale}.

Analysis of the dependence of $\Delta\sigma^{({\rm CI,CD})}_{p_t:n}$
on the upper $p_t$ acceptance cut indicates significant contribution from
particles with $p_t > 0.6$~GeV/$c$.  Subsequent studies of like-sign and
unlike-sign two-particle correlations on transverse momentum space~\cite{ptxptci} for these
data confirm that much of
the observed fluctuations result from correlation excess for $p_t > 0.6$~GeV/$c$.  The larger magnitude of unlike-sign correlations
relative to like-sign at higher $p_t > 0.6$~GeV/$c$ also results in 
$\Delta \sigma^{({\rm CD})}_{p_t:n} < 0$.  These results implicate semi-hard
scattering in the initial stage of Au-Au collisions as a possible
mechanism contributing to $\Delta \sigma^{({\rm CI})}_{p_t:n}$ and
$\Delta \sigma^{({\rm CD})}_{p_t:n}$. Strong dependence of $F_{p_t}$ on the upper $p_t$ acceptance was also reported by the PHENIX experiment~\cite{phenix200}. It is therefore of interest to examine the predictions of available theoretical collision models which include hard parton scattering and/or hadronic rescattering.

\section{Model Predictions}
\label{Section:Theory}

{\sc hijing}~\cite{jetquench}, which incorporates p-p soft scattering and longitudinal color-string fragmentation phenomenology plus hard parton scattering and fragmentation coupled to a Glauber model of A-A collision geometry, predicts a range of $\Delta\sigma^{({\rm CI})}_{p_t:n}$ up to only one-half the observed values in Fig.~\ref{minbias}. {\sc hijing} predictions include: 1) jet production enabled but without jet quenching (produces maximum fluctuations but still only one-half the measured values); 2) jet production and jet quenching both enabled (variance excess reduced by about half); and 3) no jet production (even smaller magnitude).  In addition to underpredicting $\Delta\sigma^{({\rm CI})}_{p_t:n}$ magnitudes, {\sc hijing} also does not reproduce the observed strong centrality dependence of the data or the non-monotonic behavior for the more central collisions, but is instead approximately independent of centrality. 


Other collision models differ in their treatment of lower $p_t$ (soft) particle production, rescattering and resonances, but do not include semi-hard parton scattering. {\sc rqmd}~\cite{rqmd} without hadronic rescattering predicts that $\Delta\sigma^{({\rm CI})}_{p_t:n}$ increases monotonically with centrality, reaching only half the observed value for central RHIC collisions.  Initial studies of scale-dependence indicate that the main contribution in the {\sc rqmd} model is from resonance decays and not minijets as for {\sc hijing}.  $\Phi_{p_t}$ predictions from {\sc u}r{\sc qmd} for Au-Au collisions at $\sqrt{s_{NN}}$ = 200~GeV~\cite{urqmd} indicate results similar to {\sc rqmd} and also reveal strong reduction of $\Phi_{p_t}$ when hadronic rescattering is included.  {\sc rqmd} and {\sc u}r{\sc qmd} predictions for $\Delta\sigma^{({\rm CI})}_{p_t:n}$ without hadronic rescattering constitute the upper limit for those models. The quark-gluon string model ({\sc qgsm}) for Pb-Pb central collisions at $\sqrt{s_{NN}}$ = 200~GeV, when linearly extrapolated to the STAR acceptance scale, predicts $\Delta\sigma^{({\rm CI})}_{p_t:n} \sim 10$~MeV/$c$~\cite{qgsm}, which is significantly less than the STAR measurement.

\section{Discussion}
\label{Section:Disc}

These fluctuation measurements, restricted to hadrons at lower $p_t$ ($<2$~GeV/$c$), indicate that even central Au-Au collisions at RHIC are not fully equilibrated because $\Delta\sigma^{({\rm CI,CD})}_{p_t:n}$ would vanish for ensembles of fully equilibrated events (except for the relatively small contributions from quantum and Coulomb correlations and resonance decays).  Instead, Au-Au collision events at RHIC remain highly structured, with respect to nonstatistical $\langle p_t \rangle$ fluctuations, as evidenced by the strong dependence on the upper $p_t$ acceptance. This result conflicts with assumptions underlying hydrodynamic and statistical (thermal) models conventionally applied to RHIC collisions.  We observe no evidence of critical fluctuations associated with a possible phase transition.  The quantity $\Delta\sigma^{({\rm CI})}_{p_t:n}$ used in this analysis quantifies the substantial differences between Au-Au collisions and simple models based on independent superposition of p-p collisions.  We have demonstrated that the observed charge-independent and charge-dependent nonstatistical
fluctuations cannot be explained in terms of final-state quantum
and Coulomb correlations and resonance decays or in terms of experimental
effects such as two-track inefficiencies and time dependences of experimental apparatus.

The observed strong energy dependence of $\Delta\sigma^{({\rm CI})}_{p_t:n}$ from SPS to RHIC and the failure of conventional theoretical models to describe these new RHIC fluctuation data indicate that significant new dynamical mechanisms play a role in Au-Au collisions at RHIC, mechanisms which substantially affect the correlation structure of final-state transverse momentum. The increase of $\Delta\sigma^{({\rm CI})}_{p_t:n}$ with $p_t$ upper limit, combined with apparent saturation and even reduction of $\Delta\sigma^{({\rm CI})}_{p_t:n}$ for the more central Au-Au collisions, suggests that semi-hard parton scattering and subsequent dissipation of parton momentum by coupling to an increasingly dense, possibly colored medium may account for these observations. Detailed studies of correlation structure in both transverse and longitudinal momentum components will be reported in the near future~\cite{ptxptci,axialcd,axialci}.



\section{Conclusions}
\label{Section:Con}

This first large-acceptance measurement of $\langle p_t \rangle$
fluctuations at RHIC reveals intriguing deviations from a statistical reference. We observe a $13.7 \pm 1.4$\%~(stat+syst) $r.m.s.$ fractional excess of charge-independent fluctuations in $\sqrt{n}\, (\langle p_t \rangle - \hat p_t)$ ($17 \pm 2$\%~(stat+syst) when extrapolated to 100\% of primary charged hadrons in the STAR acceptance) for the 15\% most-central events which varies smoothly and non-monotonically with centrality. This observation of strong nonstatistical $\langle p_t \rangle$ fluctuations demonstrates that RHIC events are not fully equilibrated, even in the lower $p_t$ sector for central events, contradicting a basic assumption of hydrodynamic and statistical models.  There is no significant evidence for anomalous event classes as might be expected from critical fluctuations. Comparisons with SPS experiments indicate that charge-independent fluctuations are qualitatively larger at RHIC, whereas charge-dependent fluctuations are not. A PHENIX result at 130~GeV for charge-independent fluctuations, compatible with zero within their systematic error, is consistent with a significant non-zero STAR measurement restricted to the PHENIX acceptance. Based upon studies of the higher $p_t$ contribution and various model comparisons we speculate that these fluctuations may be a consequence of semi-hard initial-state scattering (minijets) followed by parton cascading in the early stage of the Au-Au collision which is not fully equilibrated  prior to kinetic decoupling~\cite{newref}. Such strong fluctuations have not been observed previously in heavy ion collisions and are at present unexplained by theory, thus pointing to the possibility of new,
or perhaps unexpected dynamical processes occuring at RHIC.  Identification
of the dynamical source(s) of these nonstatistical fluctuations is 
underway~\cite{ptxptci} and will continue to be studied in the future.


\vspace{0.2in}
\centerline{\bf ACKNOWLEDGEMENTS}
\vspace{0.1in}

We thank the RHIC Operations Group and RCF at BNL, and the
NERSC Center at LBNL for their support. This work was supported
in part by the HENP Divisions of the Office of Science of the U.S.
DOE; the U.S. NSF; the BMBF of Germany; IN2P3, RA, RPL, and
EMN of France; EPSRC of the United Kingdom; FAPESP of Brazil;
the Russian Ministry of Science and Technology; the Ministry of
Education and the NNSFC of China; IRP and GA of the Czech Republic,
FOM of the Netherlands, DAE, DST, and CSIR of the Government
of India; Swiss NSF; the Polish State Committee for Scientific 
Research; and the STAA of Slovakia.

\appendix
\section{ }


In this appendix the {\em total variance} is defined. The {\em scale invariance} of total variance, an alternative statement of the central limit theorem~\cite{handbook,cltps}, then motivates the definition of fluctuation measure $\Delta\sigma^{2({\rm CI})}_{p_t:n}$ used in this analysis. 

A detector acceptance $(\Delta \eta,\Delta \phi)$ (generically $\Delta x$) on axial momentum space ($\eta,\phi$) can be divided into bins of size $(\delta \eta,\delta \phi)$ (generically $\delta x$). Each bin then contains event-wise scalar $p_t$ sum
\bea
\label{EqA1}
p_{t,\alpha}(\delta x) & \equiv & \sum_{i=1}^{n_{\alpha}(\delta x)}
p_{t,\alpha i},
\eea
where $\alpha$ is a bin index and $n_{\alpha}(\delta x)$
is the event-wise multiplicity in bin $\alpha$.  Fluctuations in
$p_{t,\alpha}(\delta x)$ {\em relative to} $n_{\alpha}(\delta x)$ could be
measured by the variance of the ratio 
$\langle p_t \rangle_{\alpha} = p_{t,\alpha}(\delta x) / n_{\alpha}(\delta x)$. However, to minimize contributions from event-wise and {\em bin-wise}
variations in $n_{\alpha}(\delta x)$ (a source of systematic error) we
instead compute the {\em total variance} of difference
$p_{t,\alpha}(\delta x) - n_{\alpha}(\delta x) \hat{p}_t$, defined by
\bea
 \Sigma^2_{p_t:n}(\Delta x,\delta x) \hspace{-.05in}  &\equiv& \hspace{-.15in} \overline
{\sum_{\alpha=1}^{M(\Delta x,\delta x)}
(p_{t,\alpha}(\delta x) - n_{\alpha}(\delta x)
\,\hat p_t)^2}, 
\label{EqA2}
\eea
where $M(\Delta x,\delta x)$ is the event-wise number of {\em occupied}
bins of size $\delta x$ in acceptance $\Delta x$ and the bar denotes an average over all events.  Typically $M(\Delta x,\delta x) = \Delta x/\delta x$, a constant for all events except when $\delta x \ll \Delta x$ and some bins are unoccupied. 

For the analysis described in this paper we are interested in two limits of
Eq.~(\ref{EqA2}), the acceptance scale $\delta x = \Delta x$
with $M=1$, and a `single-particle' scale $\delta x \ll \Delta x$ such that
each occupied $(\eta,\phi)$ bin contains a single particle, with $M\rightarrow n(\Delta x) \equiv N(\Delta x)$, the event-wise total multiplicity in the acceptance.  For a collection of reference events ({\em cf\,} Sec.~\ref{Section:Dist}) obtained by independent $p_t$ sampling from a fixed parent distribution (also referred to here as `CLT conditions')
quantity $\Sigma^2_{p_t:n}(\Delta x,\delta x)$ is independent of bin size $\delta x$.  We illustrate this {\em scale invariance} under CLT conditions for the above two limits and for arbitrary scale $\delta x$ as follows.

In the single-particle scale limit each occupied bin contains only one particle, and the bin index is equivalent to a particle index:
$p_{t,\alpha}(\delta x) \rightarrow p_{t,i}$ (transverse momentum of
particle $i$) and $n_{\alpha}(\delta x) \rightarrow 1$.
$\Sigma^2_{p_t:n}(\Delta x,\delta x)$ then has the limit
\bea
 \Sigma^2_{p_t:n}(\Delta x,\delta x \ll \Delta x)
& \rightarrow & \bar{N}(\Delta x) \sigma^2_{\hat{p}_t},
\label{EqA4}
\eea
where $\bar{N}(\Delta x)$ is the mean total event multiplicity, and the variance of the inclusive $p_t$ distribution is explicitly $\sigma^2_{\hat{p}_t} = \overline{\sum_{i=1}^{N(\Delta x)}
(p_{t,i} - \hat{p}_t)^2} / \bar{N}(\Delta x)$.  
In the limit $\delta x \rightarrow \Delta x$, $M(\Delta x,\delta x) \rightarrow 1$, the event-wise single-bin occupancy is $N(\Delta x)$, and $\Sigma^2_{p_t:n}(\Delta x,\delta x)$ becomes
\bea
\Sigma^2_{p_t:n}(\Delta x,\delta x = \Delta x) & = &
\sum_{N(\Delta x)} p_N \, N(\Delta x)^2 \sigma^2_{\langle p_t \rangle_N},
\label{EqA5}
\eea
where the sum includes all values of event multiplicity $N(\Delta x)$ represented in the event ensemble, $p_N \equiv \varepsilon_N/\varepsilon$ is the fraction of events in the ensemble with multiplicity $N(\Delta x)$, and $\sigma^2_{\langle p_t \rangle_N}
\equiv  \overline{(\langle p_t \rangle_N - \hat{p}_t)^2}$ is the variance of the $\langle p_t \rangle$ distribution for the subset of events with multiplicity $N(\Delta x)$.  If CLT conditions apply, then
\bea
\Sigma^2_{p_t:n}(\Delta x,\Delta x) & \stackrel{\rm (CLT)}{=} &
\sum_{N} p_N \, N(\Delta x) \sigma^2_{\hat{p}_t}
 \\ \nonumber
 & = & \bar{N}(\Delta x) \sigma^2_{\hat{p}_t},
\label{EqA5} 
\eea
where CLT relation $\sigma^2_{\langle p_t \rangle_N} = \sigma^2_{\hat{p}_t} / N(\Delta x)$ was invoked.  The equivalence under CLT conditions of $\Sigma^2_{p_t:n}(\Delta x,\delta x)$ for these two limiting scale values is thus established.

Generalizing the latter argument, the total variance at arbitrary scale $\delta x$ in Eq.~(\ref{EqA2}) can be reexpressed as
\bea
\label{EqA7}
 \Sigma^2_{p_t:n}(\Delta x,\delta x)   & = & \overline
{\sum_{\alpha=1}^{M(\Delta x,\delta x)} n^2_{\alpha}(\delta x)
(\langle p_t \rangle_{\alpha} - \hat{p}_t )^2} \\
\label{EqA8}
 & = & M(\Delta x,\delta x)\, \sum_{n(\delta x)} p_n
n^2(\delta x) \sigma^2_{\langle p_t \rangle_{n}}, \nonumber
\eea
where sums over events and bins were rearranged as sums over
bin-wise multiplicity $n(\delta x)$ and over bins $\alpha$ which have that
value of multiplicity, $p_n$ is the fraction of bins in the event ensemble with multiplicity $n(\delta x)$, and
$\sigma^2_{\langle p_t \rangle_{n}}$ is the variance of
$\langle p_t \rangle - \hat{p}_t$ within that subset of bins
\bea \label{EqA9}
\sigma^2_{\langle p_t \rangle_{n}} & \equiv & \overline{
(\langle p_t \rangle_{n} - \hat{p}_t )^2 }.
\eea
The bar in Eq.~(\ref{EqA9}) indicates an average over all bins in the event ensemble with multplicity $n$.  For CLT conditions $ \sigma^2_{\langle p_t \rangle_{n}} = \sigma^2_{\hat{p}_t} / n(\delta x)$ for any $n$,
 and, since $M(\Delta x,\delta x)\, \bar n(\delta x) = \bar N(\Delta x)$, Eq.~(\ref{EqA7}) therefore becomes
\bea
\Sigma^2_{p_t:n}(\Delta x,\delta x)   & = & \bar{N}(\Delta x) \sigma^2_{\hat{p}_t}, 
\eea
which demonstrates the general scale invariance of $\Sigma^2_{p_t:n}(\Delta x,\delta x)$ for CLT conditions. 

Deviations from central limit conditions signal the presence of two-particle correlations ({\em e.g.,} $p_t$ samples are not independent). The total variance is then no longer scale invariant, and its scale dependence reflects the detailed structure of those correlations. We therefore define a {\em total variance difference} between arbitrary scales $\delta x_1$ and
$\delta x_2$, where $\delta x_1 < \delta x_2$, as
\bea
 & &  \Delta\Sigma^2_{p_t:n}(\Delta x, \delta x_1, \delta x_2) 
\nonumber \\
  &\equiv& \Sigma^2_{p_t:n}(\Delta x, \delta x_2)\!-\!\Sigma^2_{p_t:n}(\Delta x, \delta x_1),
\label{EqA3}
\eea
\vspace{0.0in}

\noindent
where $\Delta\Sigma^2_{p_t:n}(\Delta x, \delta x_1, \delta x_2)=0$ if CLT conditions apply in the scale interval $[\delta x_1,\delta x_2]$.

The total variance difference depends by construction on the detector acceptance (and on the collision system or particpant number). We can remove those dependences in several ways whose choice depends on the physical mechanisms producing the correlations. For this application we divide by the total multiplicity in the acceptance to obtain a fluctuation measure {\em per final-state particle}.

If CLT conditions are approximately valid, $n(\delta x) \sigma^2_{\langle p_t \rangle_n}$ in Eq.~(\ref{EqA7}) is nearly constant and can be removed from the weighted summation over $n$, resulting in
\bea
\Sigma^2_{p_t:n}(\Delta x,\delta x)
 & \simeq &
\bar{N}(\Delta x) \, \overline{
n(\delta x)
(\langle p_t \rangle - \hat{p}_t )^2 },
\label{EqA10}
\eea
a factorized form in which acceptance and scale dependences are separated.
The total variance difference for $\delta x_2 = \delta x$ and $\delta x_1
\ll \Delta x$ is then given by
\bea  \label{EqA11}
 & &  \Delta\Sigma^2_{p_t:n}(\Delta x,\delta x_1 \ll \Delta x, \delta x)
\nonumber \\
 & ~~~~~\simeq &  \bar{N}(\Delta x)
  \times \left[ \overline{ n(\delta x)(\langle p_t \rangle - \hat{p}_t )^2 }
- \sigma^2_{\hat{p}_t} \right]    \nonumber \\
 & ~~~~~\equiv & \bar{N}(\Delta x)\, \Delta \sigma^2_{p_t:n}(\delta x) 
\eea
combining Eqs.~(\ref{EqA4}) and (\ref{EqA10}).  In Eqs.~(\ref{EqA10}) and (\ref{EqA11}) the bar denotes an event-wise sum over occupied bins and an average over all events.  The $\langle p_t \rangle$
fluctuation excess measure $\Delta \sigma^{2({\rm CI})}_{p_t:n}$ in Eq.(\ref{Eq3}) is thus identified as the total variance difference in Eq.~(\ref{EqA11}) {\em per final-state particle}, evaluated at the acceptance scale $\delta x = \Delta x$. 



\begin{thebibliography}{99}
\bibitem{endpoint}
$N_0$ is the half-max point at the end of the minimum-bias
distribution plotted as $d\sigma / dN_{ch}^{1/4}$, 
and is an estimator on $N$ for the maximum number of participant nucleons,
in which case
$N_{part}/N_{part,max} \simeq N/N_0$ within 4\%.
\bibitem{footnote-scale}
The assumption underlying this naive scaling is that the correlation range
is much greater than the scale, such that $\Delta \sigma^{({\rm CI})}_{p_t:n}$
is proportional to $\delta\eta$ and $\delta\phi$.
\end{thebibliography}
\end{document}